\documentclass[Physsubmission, Phys]{SciPost}

\hypersetup{
    colorlinks,
    linkcolor={red!50!black},
    citecolor={blue!50!black},
    urlcolor={blue!80!black}
}

\usepackage[bitstream-charter]{mathdesign}
\DeclareSymbolFont{usualmathcal}{OMS}{cmsy}{m}{n}
\DeclareSymbolFontAlphabet{\mathcal}{usualmathcal}

\begin{document}

\begin{center}{\Large \textbf{
Status of DMRadio-50L and DMRadio-m$^3$\\
}}\end{center}

\begin{center}
Nicholas M. Rapidis\textsuperscript{1$\star$}\\
on behalf of the DMRadio Collaboration
\end{center}

\begin{center}
{\bf 1} Department of Physics, Stanford University, Stanford, CA 94305, USA
\\
* rapidis@stanford.edu
\end{center}

\begin{center}
\today
\end{center}


\definecolor{palegray}{gray}{0.95}
\begin{center}
\colorbox{palegray}{
  \begin{tabular}{rr}
  \begin{minipage}{0.1\textwidth}
    \includegraphics[width=30mm]{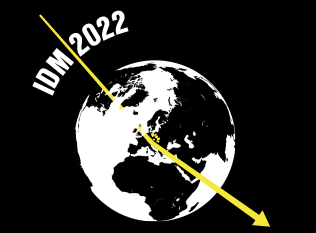}
  \end{minipage}
  &
  \begin{minipage}{0.85\textwidth}
    \begin{center}
    {\it 14th International Conference on Identification of Dark Matter}\\
    {\it Vienna, Austria, 18-22 July 2022} \\
    \doi{10.21468/SciPostPhysProc.?}\\
    \end{center}
  \end{minipage}
\end{tabular}
}
\end{center}

\section*{Abstract}
{\bf
Recent theoretical advancements have made the QCD axion a stronger dark matter candidate, especially in the sub-$\mu$eV range. While cavity haloscopes have made significant progress in excluding QCD axions in the $1-100\ \mu\text{eV} $ region, the $1 \text{ peV}- 1\ \mu\text{eV}$ region remains unexplored. The DMRadio program consists of a series of experiments designed to probe low mass axions. DMRadio-50L uses a 1 T average field toroidal magnet and a high-Q LC-oscillator with target sensitivity to axions of $g_{a\gamma\gamma} < 5 \times 10^{-15}\text{ GeV}^{-1}$between 5 kHz and 5 MHz. DMRadio-m$^3$ consists of a higher frequency LC-oscillator in a 4 T peak field solenoidal magnet with sensitivity to the DFSZ model of QCD axions between 30 MHz and 200 MHz. In this work, we present the status of DMRadio-50L and DMRadio-m$^3.$
}

\vspace{10pt}
\noindent\rule{\textwidth}{1pt}
\tableofcontents\thispagestyle{fancy}
\noindent\rule{\textwidth}{1pt}
\vspace{10pt}

\section{Introduction}
\label{sec:intro}
The QCD axion is a  well motivated dark matter candidate as it can resolve the Strong CP problem of quantum chromodynamics (QCD) \cite{Peccei:1977hh,Peccei:1977ur,Wilczek:1977pj,Weinberg1977} while simultaneously having a favorable production mechanism which can populate the universe with the currently observed abundance of cold dark matter (DM) \cite{Abbott:1982af,Preskill1983,Dine1983}. In particular, through interactions with QCD, the axion obtains a mass, $m_a \approx 5.7\text{ neV}(10^{15}\text{ GeV}/f_a )$, where $f_a$ is the energy scale associated with the breaking of the Peccei-Quinn symmetry \cite{Borsanyi:2016ksw}. Recent theoretical results have shown that the axion can be a well motivated dark matter candidate if this symmetry breaking scale occurs above the energy scale of inflation \cite{Graham:2018jyp,Tegmark:2005dy}. Since $f_a$ could be as high as the Planck scale, this has motivated experimental efforts to search for low mass (peV-$\mu$eV) axions.


Many experimental searches for the axion utilize its coupling to the Standard Model through a term in the Lagrangian of the form
\begin{equation}\mathcal{L}\supset g_{a\gamma\gamma}aF_{\mu\nu}\tilde{F}^{\mu\nu}=\frac{1}{4}g_{a\gamma\gamma}a\mathbf{E}\cdot\mathbf{B}\label{eq:lagrangian}\end{equation} where $g_{a\gamma\gamma}$ is the axion-photon coupling strength that scales inversely with the symmetry breaking scale $f_a$, $a$ is the axion field, and $F_{\mu\nu}$ is the electromagnetic field tensor \cite{Sikivie1983}. Through this interaction, an axion in a magnetic field produces a real photon whose frequency matches the mass of the axion ($\nu=m_ac^2/h)$. This is the interaction that cavity haloscopes have used to exclude portions of the QCD axion parameter space \cite{Hagmann1990,Asztalos2001,PhysRevLett.124.101303,Brubaker2017,Backes2021,PhysRevD.103.102004,PhysRevLett.128.241805}.

\section{LC-resonators as axion detectors}
The axion-photon coupling shown in Equation \ref{eq:lagrangian} leads to a modification of Maxwell's equations by introducing axion-field dependent terms. In the presence of a DC magnetic field, the axion can be modeled as an effective oscillatory current density:
\begin{equation}
\mathbf{J}_\text{eff}\approx g_{a\gamma\gamma}\sqrt{2\rho_\text{DM}}\cos(m_at)\mathbf{B}
\label{eq:cur}
\end{equation}
where $\rho_\text{DM}\approx 0.45\text{ GeV}/\text{cm}^3$ is the local DM density \cite{de_Salas_2021}. This is an accurate formalism since the number density of axions per quantum state is much greater than unity and hence they can be modeled as a classical wave.

To couple to this effective current, experiments can utilize microwave cavities in a magnetic field that resonate when the frequency of the axion current matches that of a cavity mode of interest \cite{Hagmann1990,Asztalos2001,PhysRevLett.124.101303,Brubaker2017,Backes2021,PhysRevD.103.102004,PhysRevLett.128.241805}. Using cavities, however, is unfavorable at lower frequencies since this would require increasingly larger cavities. As such, the DMRadio program uses resonators in the lumped-element regime where the resonance is the LC-resonance of the device, and thus the corresponding wavelength of the signal is much larger than any characteristic length scale of the detector. 

The operating principle of DMRadio \cite{PhysRevD.92.075012,DMRadio_Pathfinder,Brouwer:2022bwo,DMRadio-50L,10.1007/978-3-030-43761-9_16}, and LC-resonator axion detectors in general \cite{Cabrera2008,PhysRevLett.112.131301,PhysRevLett.117.141801,PhysRevLett.122.121802,PhysRevD.99.055010,PhysRevLett.127.081801,Gramolin2020a,PhysRevLett.124.241101,PhysRevLett.126.041301}, is that a magnetic field is placed inside an inductive element. The axion then produces the effective current  (Equation \ref{eq:cur}) in this inductor and the resulting screening currents on the inductor walls can ultimately be measured by a device such as a DC SQUID. By placing a capacitor in series with the inductor, an LC-resonance is achieved, with a quality factor $Q$, and hence the current is enhanced by a factor of $Q$ on resonance. 





\section{DMRadio-50L}
\begin{figure}[h]
\centering
\includegraphics[width=0.72\textwidth]{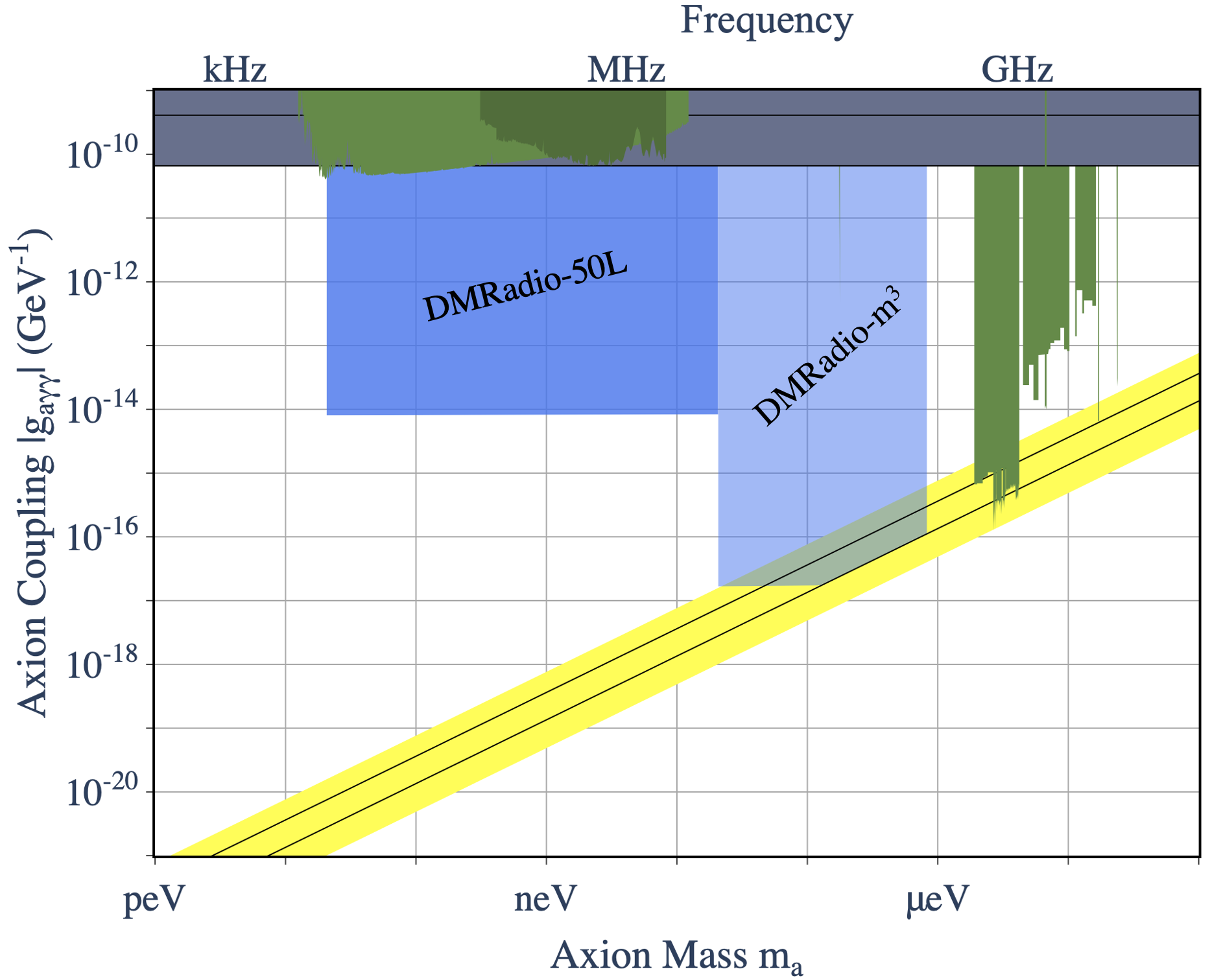}
\caption{The anticipated exclusion limits for DMRadio-50L (dark blue) and DMRadio-m$^3$ (light blue). The QCD axion band is shown in yellow. Existing limits from CAST \cite{CAST:2017uph} are shown in dark gray and haloscope exclusions are shown in green. DMRadio-50L has sensitivity to axions with mass $20\text{ peV}<m_a<20\text{ neV}$ (5 kHz $<\nu_a<$5 MHz) at an axion-photon coupling of $g_{a\gamma\gamma}>5\times10^{-15}\text{ GeV}^{-1}$ and DMRadio-m$^3$ has sensitivity to  DFSZ axions at $100\text{ neV}<m_a<800\text{ neV}$ ($30\text{ MHz}<\nu_a<200\text{ MHz}$) and KSVZ axions at $40\text{ neV}<m_a<100\text{ neV}$ (10 MHz $<\nu_a<$ 30 MHz). Following the conventions of the community, all these limits are set using the standard halo model, however other models are being considered by the DMRadio collaboration.}
\label{fig:exclusion}
\end{figure}

DMRadio-50L excludes axions at an axion-photon coupling of $g_{a\gamma\gamma}>5\times10^{-15}\text{ GeV}^{-1}$ in the mass range of $20\text{ peV}<m_a<20\text{ neV}$ (5 kHz $ <\nu_a<5$ MHz), shown in Figure \ref{fig:exclusion}. Apart from its sensitivity to axions, the DMRadio-50L experiment acts as a testbed platform for novel quantum sensors. This experiment is being built on the Stanford University campus.

\begin{figure}[h]
\centering
\includegraphics[width=0.94\textwidth]{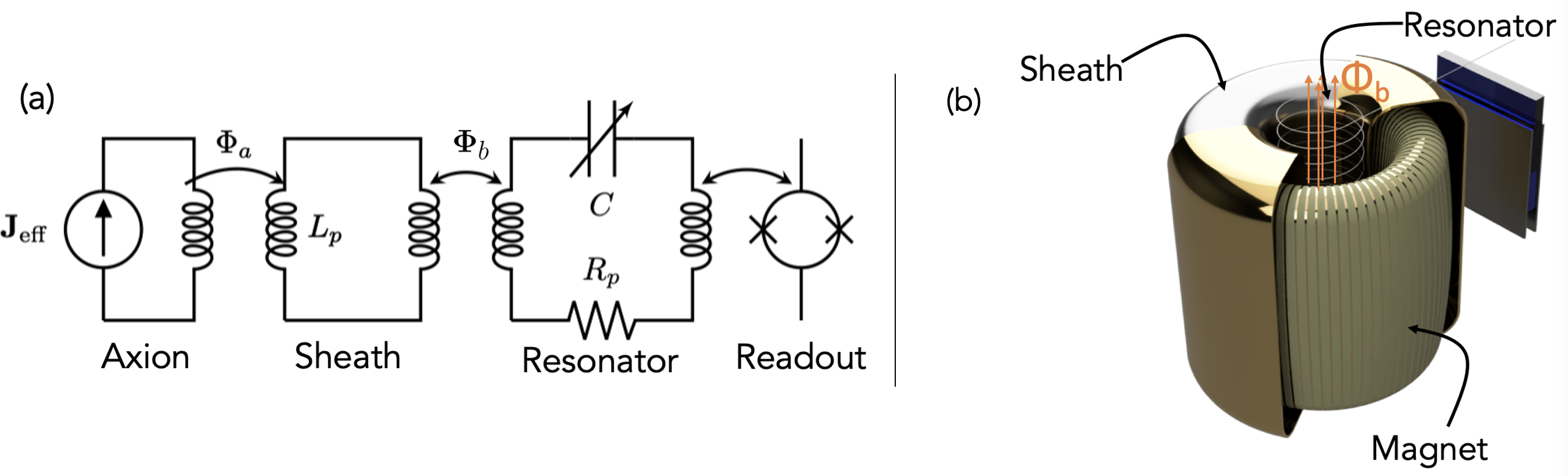}
\caption{(a) A full circuit diagram for the layout of DMRadio-50L. The effective axion current inductively couples to the sheath which then couples to the inductor in the center of the toroid. The LC resonator is then inductively coupled to a readout device like a DC SQUID. (b) An illustration of key components of DMRadio-50L. Part of the sheath has been hidden to reveal the magnet inside it. The flux $\Phi_b$ is the axion induced flux due to the screening currents on the outside of the sheath.}
\label{fig:design}
\end{figure}

The experiment utilizes a $\sim50$ L toroidal magnet that sustains an average DC magnetic field of 1 T. A superconducting sheath is then wrapped around the magnet. Since the effective axion current flows within the volume of the magnet, it inductively couples to the sheath, thus producing screening currents on the inner walls of the sheath. By cutting vertical slits in the toroidal sheath, the currents flow from the inner walls of the sheath onto the outer walls -- this ultimately induces an AC flux in the central hole of the toroid which is  picked up by an inductor in an LC resonator. The capacitive portion of the resonator is tunable, which allows for DMRadio-50L to resonantly scan over several decades in mass. The circuit diagram for the experiment as well as a corresponding illustration of the detector are shown in Figure \ref{fig:design}.

To avoid having the inductor couple to the lossy components of the magnet within the sheath, the slits are covered by superconducting elements whose minor radius is larger than that of the sheath.

All of these components are placed in a superconducting shield and then cooled using a dilution refrigerator. To minimize thermal noise, the resonator is held at a base temperature of approximately 20 mK. As of Summer 2022, all major components of DMRadio-50L have either been delivered to the site location or are under construction, and final assembly of the experiment will occur in Summer 2023.

While the first commissioning of the experiment employs DC SQUIDs for the readout, future iterations of the experiment will employ and test amplifiers with added noise below the standard quantum limit (SQL) \cite{Kuenstner:2022,Kuenstner2:2022}. Since the sensitivity of DMRadio is set by the sensitivity bandwidth of the experiment \cite{Chaudhuri:2018rqn}, and not on-resonance behavior, one can benefit from techniques that take advantage of the tradeoffs between the imprecision noise and backaction noise of quantum amplifiers to achieve a significant increase in scan rate. 


\section{DMRadio-m$^3$}

DMRadio-m$^3$ has sensitivity to DFSZ axions \cite{Dine:1981rt,Zhitnitsky:1980tq} at $100\text{ neV}<m_a<800\text{ neV}$ (30 MHz $<\nu_a<$ 200 MHz) and KSVZ axions \cite{SHIFMAN1980493,PhysRevLett.43.103} at $40\text{ neV}<m_a<100\text{ neV}$ (10 MHz $<\nu_a<$30 MHz) as shown in Figure \ref{fig:exclusion} \cite{Brouwer:2022m3}. This experiment is being built at SLAC National Lab.

Since the resonance frequencies of DMRadio-m$^3$ are higher than those of DMRadio-50L, a toroidal geometry would have parasitic resonances that would diminish the sensitivity to axions in portions of the desired frequency range. As such, the optimal design for such a structure utilizes a solenoidal magnet -- this magnet sustains a peak DC field of $>4$ T.  To couple the vertically oscillating axion current, a copper coaxial pickup structure is placed inside the magnetic field. Tuning of the resonance frequency is achieved by placing a tunable capacitor across this inductive coax. 

Bucking coils are incorporated to steeply reduce the magnetic field above the coaxial structure such that superconducting elements can be placed there. These elements include the tunable capacitor for the LC-resonator and the DC SQUIDs. As such, the primary loss source for the circuit is set by the electron losses in the non-superconducting copper coax.

The design of DMRadio-m$^3$ is currently being completed as one of the six experiments funded under the US Department of Energy Dark Matter New Initiatives program.

\section{Conclusion}
DMRadio-50L and DMRadio-m$^3$ are poised to exclude axions in the peV-$\mu$eV region with world-leading sensitivity. Alongside future experiments such as DMRadio-GUT \cite{Brouwer:2022bwo} which utilizes high-field and high-volume magnets as well as beyond-SQL amplifiers, the DMRadio program covers a significant portion of the sub-$\mu$eV axion parameter space at DFSZ sensitivity. Recent theoretical advancements have made the axion an even more attractive candidate dark matter candidate, especially at low masses. 

This series of experiments will also act as a platform for testing novel beyond-SQL amplifiers operating in the thermal limit. Upon employing these amplifiers, such experiments gain a significant enhancement in the scan rate \cite{Chaudhuri:2018rqn}.

\section*{Acknowledgments}
This research is funded in part by the Gordon and Betty Moore Foundation. Additional support was provided by the Heising-Simons Foundation. The authors also acknowledge support for DMRadio-m$^3$ as part of the DOE Dark Matter New Initiatives program under SLAC FWP 100559. Members of the DMRadio Collaboration acknowledge support from the NSF under awards 2110720 and 2014215. Part of this work was performed in the nano@Stanford labs, which are supported by the National Science Foundation as part of the National Nanotechnology Coordinated Infrastructure under award ECCS-1542152. S.~Chaudhuri acknowledges support from the R.~H.~Dicke Postdoctoral Fellowship and Dave Wilkinson Fund at Princeton University. Y.~Kahn was supported in part by DOE grant DE-SC0015655.  B.~R.~Safdi  was  supported  in part  by  the  DOE  Early  Career  Grant  DESC0019225. P.~W.~Graham acknowledges support from the Simons Investigator Award no. 824870 and the Gordon and Betty Moore Foundation Grant no. 7946. J.~W.~Foster was supported by a Pappalardo Fellowship.

\bibliography{IDMRapidis_bib.bib}

\nolinenumbers

\end{document}